# TACTICS: TACTICal Service Oriented Architecture[1]

Alessandro Aloisio and Marco Autili, *University of L'Aquila, Italy*
Alfredo D'Angelo, *Thales Italia*
Antti Viidanoja, *Patria, Finland*
Jérémie Leguay, *Thales Communications & Security, France*
Tobias Ginzler, *Fraunhofer FKIE, Germany*

Thorsten Lampe, *Thales Defence & Security, Germany*
Luca Spagnolo, *Selex ES, Italy*
Stephen Wolthusen, *Gjøvik University College, Norway*
Adam Flizikowski, *ITTI, Poland*
Joanna Śliwa, *Military Communication Institute, Poland*

**Abstract**

*Due to the increasing complexity and heterogeneity of contemporary Command, Control, Communications, Computers, & Intelligence systems at all levels within military organizations, the adoption of the Service Oriented Architectures (SOA) principles and concepts is becoming essential.*

*SOA provides flexibility and interoperability of services enabling the realization of efficient and modular information infrastructure for command and control systems. However, within a tactical domain, the presence of potentially highly mobile actors equipped with constrained communications media (i.e., unreliable radio networks with limited bandwidth) limits the applicability of traditional SOA technologies.*

*The TACTICS project aims at the definition and experimental demonstration of a Tactical Services Infrastructure enabling tactical radio networks (without any modifications of the radio part of those networks) to participate in SOA infrastructures and provide, as well as consume, services to and from the strategic domain independently of the user's location.*

## 1. Introduction

Nowadays, information exchange and integration with Command & Control (C2) systems [1,2], and Command, Control, Communications, Computers & Intelligence (C4I) systems [3] is feasible only where the available communication infrastructure physically permits message transfer (usually online HTTP/TCP/IP access, E-Mail, MIP over TCP sessions) with sufficient bandwidth and availability. However, for land based military operations this kind of connectivity can only be assumed from brigade level upwards and, probably, for a limited number of specifically equipped vehicles.

In this setting, the Network Enabled Capability (NEC) concept [4,5] is being introduced as the coherent integration of sensors, decision-makers, effectors and support capabilities to achieve more flexibility and responsiveness. It requires tactical processes to be increasingly integrated in the services and information infrastructures of the strategic domain (e.g., blue force tracking). The problem here is that the underlying technology is not reflecting this shift in NEC operations. That is, NEC requirements pose particular challenges and require the development of new technologies to maximize the military benefit. A SOA based approach, accessible in and from the tactical domain, can be the answer to these challenging requirements.

A key objective is to improve situational awareness and reactivity, i.e., the capability to quickly benefit from new information that changes the tactical situation. Usually, this depends on a complete reaction process, involving environmental data, data services, and decision makers, all of them interconnected. In the context of army Geographical Information Systems (GIS) and especially the infantry war fighter, information comes from sensors such as radar, cameras, PDAs, touch pads, etc. This domain is characterized by potentially highly mobile actors equipped with constrained communications media (unreliable radio networks with limited bandwidth) which limit the applicability of traditional SOA technologies for the tactical domain.

The TACTICS project is addressing this problem through the definition and experimental demonstration of a Tactical Services Infrastructure (TSI) enabling tactical radio networks (without any modifications of the radio part of those networks) to participate in SOA infrastructures and provide, as well as consume, services to and from the strategic domain independently of the user's location. The TSI provides efficient information transport to and from the tactical domain, applies appropriate security mechanisms, and develops robust disruption- as well as delay-tolerant schemes. This includes the identification of essential services for providing both a basic (core) service infrastructure and enabling useful operational (functional) services at the application level.

An interesting goal of the project is to provide an experimental setup (including a complete network setup using real radio communication, and running exemplary applications and operational services) to demonstrate the feasibility of the concepts in a real-life military scenario. To summarize, the final goals of the TACTICS project are to:

- propose the definition of a service-oriented architecture (SOA) compatible with the constraints of tactical radio networks;
- suggest feasible ways of adapting services to the constraints of the tactical radio networks;
- demonstrate the capacity of a TSI to offer operational services in a real tactical environment.

The paper is structured as follows. Section 2 delineates the problem space; Section 3 reports on our analysis of the TACTICS domain; Section 4 describes the solution proposed by TACTICS and Section 5 discusses the targeted progress beyond the state of the art. Section 6 concludes the paper and outline future directions.

## 2. Problem Space

The identification of new military capabilities, which most closely correspond to the military user's needs, is one of the most important expected outcome of the TACTICS project. To this aim, Section 2.1 describes some of the desired capabilities that are expected in a tactical environment; Section 2.2 describes the technical requirements that are to be fulfilled in order to achieve the described military capabilities; Section 2.3 concludes this section by summarizing the key challenges that must be addressed.

### 2.1. New Military Capabilities

To the purposes of the TACTICS project, a military capability is defined as the ability to achieve a desired effect under specific conditions by combining ways and means to perform a task. The EU Network Enabled Capability Implementation Study [7,8,9] specifies a set of desired capabilities that are in the main focus of the project.

Practically, in a tactical environment, every single combat system will be or will provide a mobile platform. Thus, some of the desired capabilities in a tactical environment are:

- Maintenance of high efficiency of operation in complex, disadvantaged and agile environment;
- Creation and management of an accurate and timely shared situational awareness among tactical platforms, which can be tailored to the specific mission needs, for use by any party involved in an operation.

### 2.2. Enabling Technical Requirements

In order to enable the military capabilities described in Section 2.1, the tactical system must be able to technically cover the following requirements:

- Interoperability of various systems and devices at the tactical level, such as sensors, effectors, C2 and C4I systems, where the participating systems may be from different nations and/or vendors;
- Establishment of SOA-centric common core services;
- Adaptation of the web-ubiquitous services paradigm to the tactical theatre;
- Enabling service discovery and dynamic composition in resource-constrained networks.

In particular, to assure the system robustness, TACTICS will deal with, real time event notifications, situational awareness, alerts and chat, service prioritization, and bandwidth pre-emption (mission-aware system).

### 2.3. Key Challenges

In order to fulfill the above requirements, hence enabling the described capabilities in a tactical environment, the following problems have to be addressed:

- Lack of connectivity guarantee;
- Changing network topology that implies no guaranteed service delivery;
- Radio silence that implies no guaranteed data delivery.

Underlying key challenges are (i) the employment of resource-constrained mobile devices in the tactical domain, as well as their increasing heterogeneity, (ii) the adoption of error-prone, low-bandwidth communication media, and (iii) the resulting infrastructure-less nature of the tactical radio network.

With reference to Figure 1, current SOA architectures used in the Strategic/Operational Domain (mainly enterprise SOA implementations, such as Enterprise Service Buses - ESBs) have an overwhelming resource demand that would overstress the scarce resources of the radio network if adopted in the Tactical Domain.

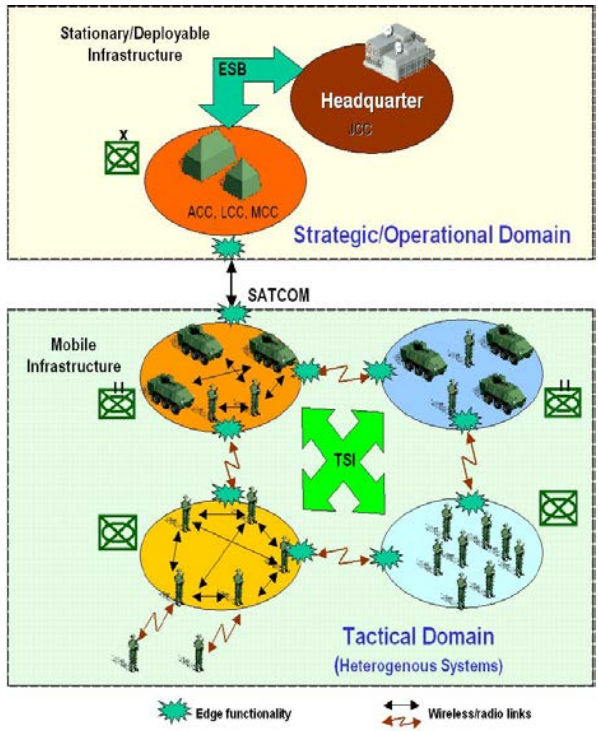

**Figure 1 Tactical Environment**

To this end, TACTICS needs to inquire the applicability of SOA traditional approaches, identify and define optimizations and enhancements that better suit the requirements of the tactical domain (e.g., being resource efficient, delay- and disruption-tolerant, enabling multicast messaging, and leverage traffic shaping methods).

## 3. TACTICS Domain Analysis

Based on the results of our analysis, a lightweight SOA stack tailored to the needs of the tactical domain have to be defined. This includes the identification of essential services for both providing a basic (core) services infrastructure and enabling useful operational (functional) services on application level.

The fragile nature of tactical radio networks requires robust communication mechanisms that current SOA solutions do not provide. This includes methods to deal with large delays, communication failures, and network splits and merges. These features together with the characteristics of low capacity devices, usually used in the tactical battlefield, are contrary to the ponderous implementation of classical SOA stacks. A smart communication infrastructure is needed to deal with these peculiarities of the wireless medium. For this purpose, an intelligent overlay network (such as a peer-to-peer network) is envisioned to enable distributed data storage/retrieval, distributed redundant service registries, and publish/subscribe mechanisms.

An important aspect of the project is to match the requirements of the services with the properties of the underlying communication technology that may involve the exchange of cross-layer information for a proper adaptation of service requirements, and a proper configuration of the network. In this context, the network Quality of Services (QoS) should be considered as an intrinsic part of the SOA layer. Special care has to be taken to ensure proper mediation in between multi-domain NEC-enabled environments with heterogeneous tactical and strategic networks as well as different data, services and user requirements.

Given the more fluid environment and changing requirements for security resulting from the application of these new models, an efficient interconnection of the tactical and strategic security domains is necessary to maintain a sufficiently secure mission environment. In particular, security mechanisms have to be defined to provide an appropriate level of trust and redundancy in the open and error-prone tactical domain.

To make TACTICS interoperable with current and future SOA solutions used in the strategic domain, the formats and procedures used in the tactical domain must be transformed and adapted to the ones used in the strategic domain. A similar kind of gateway functionality may also be required to ensure interoperability between arbitrary domain borders such as those defined by security, communication, information, and coalition related constraints. At the same time, dynamic changes in coalitions and information exchanged by partners must be supported actively by security mechanisms.

Modern coalition operations are conducted in a dynamic environment usually with unanticipated partners and irregular adversaries. In order to act successfully they need technical support that gives modularity and flexibility in connecting heterogeneous systems of cooperating allies.

## 4. Solution Space

In order support the described cooperative scenario, Service Oriented Architectures are recommended within the NATO community, through the Strategy for Developing Future Consultation and Command and Control Capabilities [6,7,8,9,10]. Indeed, also the EDA Capabilities Directorate presents a similar approach [11]. Figure 2 shows the vision and concepts of the NNEC framework, further highlighting the TACTICS focus.

According to the NNEC and EDA strategies, an important observation is that information exchange should be seen not only in terms of what derives from the usual vertical relation between the commander and the subordinate. Rather, looking down to the battalion, platoon and lower levels, operators more and more frequently exchange information horizontally, sharing information within small formations (platoon or squad), e.g., positions, alarms, video streams, pictures and other important elements of building situational awareness at the tactical level – and necessarily in real time. That is, individual soldiers are often equipped with high-quality sensors and hardware that turn them into complex technical systems on their own. Each actor engaged in an operation can thus be considered both as a consumer and as a provider of data.

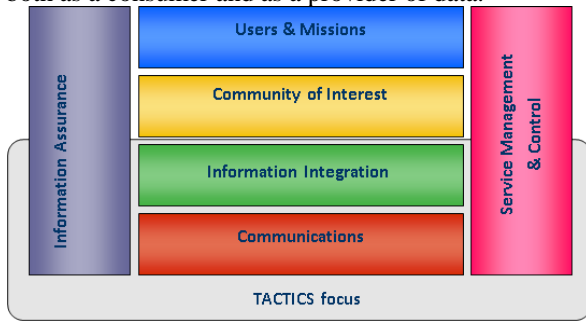

**Figure 2 TACTICS focus within NNEC**

Down the echelons of command, actions are getting more and more dynamic, making decision-making time shorter, so that actors need real-time information. Moreover, information granularity low in the military hierarchy is greater since military staff need detailed information about their area of operation.

Tactical SOA solutions will tackle these issues providing the possibility to create Common Operational Picture (COP) on various levels of command. COP created at high levels of command, tailored to the needs of the operators and giving them the overview of coalition, neutral and enemy forces, enabling to plan and conduct operation in real time can now be achieved at the lowest levels including the individual soldier. This supports creating shared situational awareness that enables military staff to make reliable decisions in a short timeframe, to work together in new, more effective ways and thereby to improve the speed of command, leading to a dramatic increase of mission effectiveness. The improvement of situational awareness and reactivity and the involvement of environment data, data services and decision makers in a tactical context are the major contributors in creating combat phases of very high intensity, concentrated within a short time framework, interleaved with more extended periods of stabilization and reconstruction.

The new concept adopted in TACTICS is focused on giving the best technology to not only protect the armed forces but also to give the individual soldier and the tactical force the complete knowledge of the operative scenario, in order to perform military action more quickly and efficiently. The existence of a flexible and well-balanced military background is the basic starting point for easy integration with allied forces. The harmonic balance and the optimization of the generalized awareness capability is the natural basis for the creation of a modern, expeditionary, net-capable and effect-based-operation-oriented common NATO force. Efforts of the EU nations have already started to go into the direction of improving technologies for situational awareness at tactical level. Examples include the ForzaNEC initiative [16] in Italy or RuDi/EiS-A-F-W in Germany [17]. However, up to now no common initiative is in progress.

TACTICS is taking the direction of filling this gap by creating a unique and homogeneous framework based on heterogeneous national tactical networks with the goal of reinforcing and effectively leveraging the ongoing separate efforts from the different nations. Even though no waveform development or adjustments will be made within the study, TACTICS will consider the deployment on different national IP radio networks, considering different national waveforms. In this analysis, both CSMA and TDMA access, as well as different characteristic waveforms, will be taken into account. Existing waveforms (20-25 km coverage, less than 100 kbit/s) will be especially considered.

Summarizing, as shown in Figure 3, the study envisions seamless end-to-end service delivery throughout the whole operational spectrum from the high operational levels down to the individual soldier in the field. To achieve this mid-term vision, TACTICS focuses on transferring the well-established SOA concepts to the tactical level in order to gainfully apply the service paradigm for:

- generating a COP on the (highly) mobile level by enabling real-time horizontal information

IV

exchange with the flexibility of service-orientation while dealing with the given bandwidth limitations;

sensors, C2 systems, effectors and support capabilities as well as radio communication.

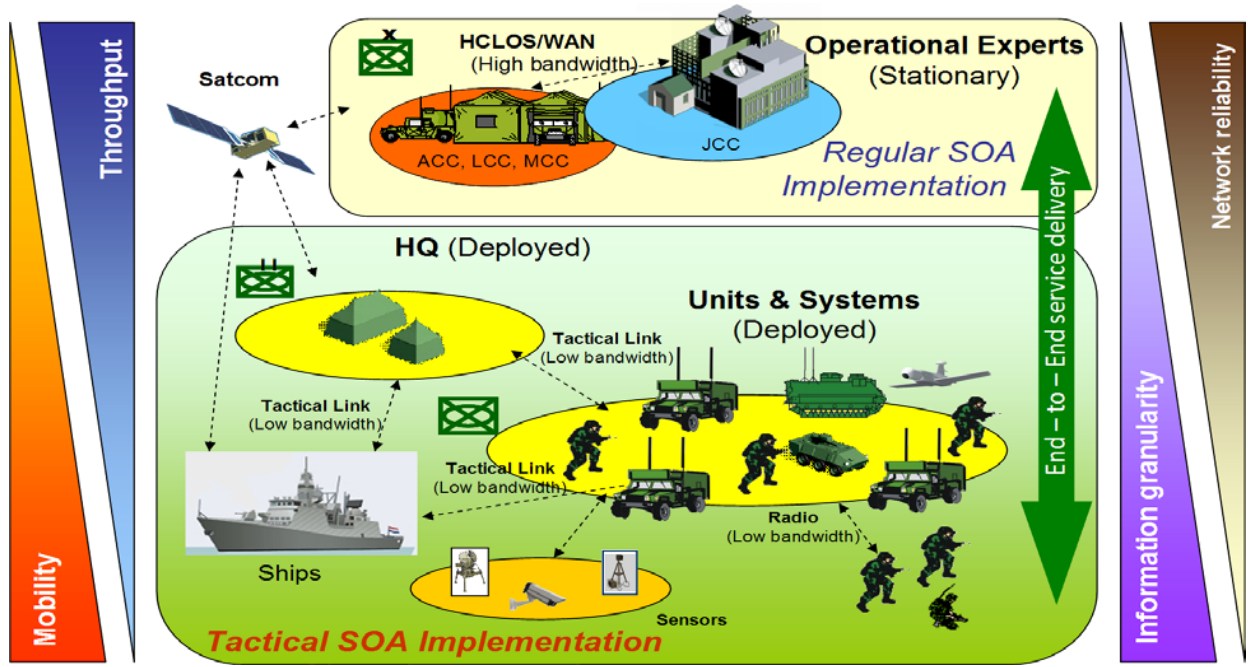

**Figure 3 TACTICS vision and beyond**

- enabling deployed tactical units to flexibly use available capabilities, switching service providers as necessary in response of radio, network or device failure, ensuring appropriate routing and service selection under the given constraints;

- guaranteeing security, policy enforcement and best possible Quality of Service (QoS) despite poor reliability of the underlying network (such as high latency or unavailability of routing nodes).

To achieve the above, TACTICS will ultimately support the NNEC and EDA Capability, Armament, Technology (CAT) Directorate vision by:

- providing concepts for integrating tactical processes into service and SOA information infrastructure, as already available for the strategic domain, thus combining regular and tactical SOA implementations;

- demonstrating the feasibility of these concepts in a real-life military scenario by building an exemplary service infrastructure that integrates existing components from the mobile and ultra-high mobile domains with one or more existing

## 5. Progress beyond the State of the Art

Specifically targeting services over constrained networks, and a lightweight implementation of the TSI capable of running on CPU- and memory-constrained servers (as often found in the field vehicles), TACTICS aims at bringing the following main improvements:

- Implementing the knowledge management of the underlying network topology so that the service bus can learn the transmission layer(s) capabilities and workload(s). The service bus is then able to find the optimal means and routes to link service producers and consumers, hence performing an optimal service routing.

- Implementing the knowledge management of which services are reachable and operative on the service bus, given that the connectivity may be lost or that the sensors/effectors presented through a given service are not operative.

- Optimizing the usage of legacy narrowband radio networks by elaborating efficient mechanisms to provide services.

- Optimizing the performance to transmit SOAP/HTTP messages with considerable

information overhead to avoid wasting the scarce radio bandwidth of the tactical networks. Correspondingly, messages must be formatted according to the operational context and situation.

- Implementing a distributed data dissemination mechanism enabling a clear decoupling between services/data producers and service consumers.

- Facilitating the deployment, configuration and management of services on the various platforms by configuring roles and functions of the platforms and actors for a given mission, rather than individually configuring the services running on each server.

- Reducing the traffic necessary for discovering service by taking into account the notion of (pre-configured) local service directories (e.g., pre-declaring a service in one geographical area).

- Enabling the specification of different service delivery levels for any given service. The assigned service level is subject to real-time negotiation with the route provider. To this end, the service specification indicates both the level of degradation and the criteria upon which this degradation takes place.

- Providing a service access control mechanism to ensure that only the authorized service consumers can access a given service provider.

- Providing a payload encryption, content integrity and signature mechanism ensuring that only those consumers with the appropriate decoding keys can access contents and check integrity, as well as the signature of the originator.

- Providing end-to-end QoS delivery services.

## 5.1 SOA Stacks & Services

In 2009, NATO has recommended a core set of services [6,7,8,9,10], including discovery, interaction, infrastructure, repository and mediation services. However, mobility and tactical requirements are not taken care of in these recommendations. As a result, the web services actually available by NATO, such as NATO Friendly Force Information[2] (NFFI), map diffusion with protocols Web Map Service[3] (WMS) and Cosumnes Community Services District[4] (CSD), all rely on high-rate networks. Moreover, the proposed test bed makes use of broadband WAN connections and focuses on interoperability only at the higher layer.

The German national initiative Referenzumgebung Dienste (RuDi [16], i.e., reference environment for services) provides a service environment down to the tactical level but, for mobile usage, requires powerful devices.

The international project Coalition Networks for Secure Information Exchange (CoNSIS [15]) considers highly mobile applications, but concentrates on improving existing ESBs such as the open source SOPERA bus, which has not yet been proven to run on smaller devices.

TACTICS explicitly targets highly mobile devices with (extremely) limited resources. It aims at specifically tailored, adaptive service stack for mobile terminals, interoperable with established SOA best practices but flexible enough to cope with the limited resources and high dynamics of a mobile tactical environment. Disruption-tolerance is expected from the service stack definition in TACTICS. In addition, semantics in the service definition and locality- and capability-aware service selection mechanisms aids the usage of services more efficiently than existing stacks. The exploitation of lightweight semantic metadata for describing services and their requirements enables the dynamic selection of the best information sources based on different criteria (location, resources consumption, time constraints), further enabling service composition at run time. Also, the possibility of having asymmetric data flow will be examined, e.g., a limited up-flow and a down-flow with a high throughput capacity. For example, this capability might support a service call on a low-bandwidth radio link (VHF/UHF) to register a video stream through a specific high-bandwidth UAV link.

## 5.2 Service Delivery

Publishing and discovering descriptions of available services are two core functions of service delivery. As briefly discussed in this section, existing solutions to service discovery cover various issues that emerged in the current Internet network. However, there is still plenty of room for research towards the new challenges introduced by the tactical networks.

We organize the discussion of service delivery in two parts: the first part concerns issues related to the way available service descriptions are organized and managed within dedicated registries, while the second part focuses on protocols for service discovery that mainly differ according to the architecture of the registry, from centralized to distributed.

---

[2] http://systematic.com/defence/products/a/command-and-control-information-systems/nffi
[3] http://www.opengeospatial.org/standards/wms
[4] http://www.yourcsd.com

For the interested reader who would like to probe further, there are certain excellent surveys worth mentioning. Specifically, in [18] the authors deeply discuss research challenges and research directions for service-oriented middleware design, therefore, investigating service description, discovery, access, and composition. In [19], the authors pose criteria for the evaluation of service registries and organize the discussion from two viewpoints: the viewpoint of the system and the viewpoint of humans. The work in [21] emphasizes the degree of distribution of the service registry and the use of semantic information in the service matchmaking process. Further works focus on service discovery protocols for mobile ad hoc environments, which are part of the Future Internet [20, 22]. All the other references to work on service delivery can be found in the above mentioned surveys.

**Service Registry -** The baseline approach concerning the organization and management of service descriptions within registries, is the data model that has been proposed in the Universal Description Discovery and Integration (UDDI) specification[5]. From the early days of service discovery, a major research issue that emerged was the enhancement of the simple UDDI data-model with semantically rich metadata. All the efforts in this line of research base their motivation on the fact that both the UDDI data-model and its querying mechanisms support keyword queries without any semantic information. In general, the semantic annotation of the service information managed by service registries requires extra specification effort. This effort may become an impediment toward the scalability of related approaches. Consequently, another research issue that arose was to incorporate in the service registries means for extracting semantically rich information out of available service descriptions in a (semi)automated way.

**Service Discovery -** Regarding service discovery protocols, the baseline approach is centralized, relying on a single registry that organizes and manages service descriptions. This provides consistency and fast local retrieval under normal circumstances. Unfortunately, the centralized approach has various drawbacks that correspond to research issues that emerged in the early days of service discovery. First, the centralized approach does not scale well with respect to the increasing number of clients that pose service discovery queries [19,21]. Moreover, in the centralized approach, the single registry constitutes a single point of failure. Therefore, the availability of centralized solutions is questionable. The aforementioned issues become even more important, in the cases where service discovery is supposed to be handled in mobile tactical environments that consist solely of mobile nodes [20,22]. To deal with the scalability, the availability and the mobility issues, several approaches proposed decentralized solutions, briefly discussed in what follows.

The distributed approaches enable peers, participating in a network of cooperating sites, to store locally their own service registry. Then service retrieval is facilitated by distributed querying services that span all these local registries to compile and present answers to users. Naturally, the distributed setting provides a richer set of answers and significant chances of better scalability in terms of both the user load and the available data to index, without the risk of a single point of failure that the centralized solution suffers from. This comes at the cost of supporting a framework that allows each peer to know which peers to contact for serving a user request along with the necessary communication overhead whenever a query to the decentralized virtual registry is posed.

The two major drawbacks for highly mobile tactical scenarios described in previous sections are the lack of a dedicated server with permanent connection to every device and the large resource demand for the server devices. It is envisioned to leverage the state of art solutions to improve the capabilities of the end user in terms of the NEC concepts described in Sections 1 to 4. The delivery methods in TACTICS are focused on bringing the right information to the right place at the right time. TACTICS aims for service discovery methods and improved registry mechanisms which truly respect the nature of tactical networks, such as the specific wireless mobile network limitations, user behavior and mobility, and constraints on military device resources. Moreover, TACTICS will investigate and develop techniques to deal with the disruptive nature of the tactical network while at the same time respecting resource efficiency.

The methods being developed in TACTICS shall conform to accepted standards (civil or military), either by modifying an existing solution, adapting it to the tactical network constraints, or by integrating new technologies.

### 5.3 End-to-end Service Management

The current research on assuring service quality in tactical wireless communications focuses on increasing bandwidth, improving reliability, and enabling adaptation by focusing on areas such as network coding, dynamic spectrum exploitation, robust routing protocols and cross-layering.

---

[5] http://uddi.xml.org

Although a variety of QoS-based routing solutions dedicated to mobile ad-hoc networks exists, the inherent unpredictability of mobile network conditions (varying from satisfactory to lack of communication) makes it impossible to guarantee quality levels throughout the overall mission. A number of approaches have been proposed in the literature to deal with QoS on military operations. These approaches include among all the Differentiated Services Architecture [23] and Advanced Resource Reservation Protocol (ARSVP) [24] for tactical networks. Civilian frameworks for combined resource management (DiffServ-based [25]) in heterogeneous IP networks to support end-to-end QoS were also studied in various European research projects (e.g., FP6 EuQoS[6], FP6 NetQoS[7]). Results of the aforementioned approaches will be the input for the process of identifying and recommending QoS and routing solutions that can be tailored to the SOA requirements within TACTICS.

Thus, there is a strong demand for middleware protocols that will benefit from combining various mechanisms available in the ISO/OSI stack (above the network layer) to tailor QoS management to the SOA stack. In addition, our plan to achieve network disruptions resiliency is to also investigate a Delay Tolerant Network (DTN) [26] component to be integrated within the middleware decision logic, when needed according to the current context.

TACTICS will define a SOA service quality management strategy and recommendations combining the status of underlying network (through monitoring), cross-layer capabilities (to improve performance provided by using only middleware-based QoS management) as well as the SOA service requirements and context (to provide optimized service delivery adapted to the network capabilities). Cross-layer quality assurance/improvement activities should consider the limitedness of tactical terminal's processing capabilities and take into account possibilities of direct and indirect gathering of particular layers' information from a terminal. As a result, the recommendations should be provided for optimal cross-layer information composition and SOA service composition/processing strategies.

The target is a holistic process that constantly plans, coordinates, negotiates, monitors and reports QoS targets to ensure that required quality is maintained or, when necessary, gradually improved. TACTICS will propose enhancements to well-recognized IT service management standards (e.g., ITIL, COBIT) to satisfy the service needs on tactical networks.

---

[6] http://www.euqos.eu
[7] http://www.netqos.eu

### 5.4 Security

TACTICS must provide a secure framework for information processing and exchange, services publishing and sharing, data model, language and protocol. The realization of the framework will be challenged by its feasibility in a tactical radio environment.

In TACTICS, there are different levels and different constraints to take into account. In particular, a multi-policy architecture will be considered, both in terms of different security properties (e.g., confidentiality) and incorporating cross-layer information (e.g to allow for restrictions on routes and services) based on a descriptive logic framework.

Towards SOA-level security, it is necessary to have a mapping between the user, the services registry and both Public Key Infrastructure (PKI) and key distribution and revocation schemes for disadvantaged networks, which allow the support of coalition environments and efficient removal of nodes and principals where compromise is suspected. This way, traffic associated with the security infrastructure can be minimised. Regarding the tactical environment, performance optimization has to be considered, both in terms of light security mechanisms or signaling optimization regarding security and on compression mechanisms.

In TACTICS, the protection of information exchange will be based on strong cryptography supporting standard and national ciphers, using an innovative key distribution methodology to ensure that information can only be accessed by authorized users. The following standards will be taken into account as a basis:

- For message level security, WS-Security, WS-Policy, and WS-Security Policy will be analyzed and investigated as initial state of the art;
- SAML and XACML are analyzed as initial state of the art for single-sign-on problems and for policy extensions;
- Complement encryption policy for "network-aware" non-data services (voice, streaming video, real time alert data);

Due to the nature of the tactical environment, specific attention will be paid to security optimization, and especially lightweight security mechanisms and signaling optimization as well as compression mechanisms are analyzed. These must be adaptable to constraints imposed by divergent node capabilities and must particularly be designed to minimize communication complexity whilst being able to recover from faults in the network substrate as well as handling compromised nodes. Correctly identifying partners in a coalition and filtering the information flow, possibly also coarsening information, are novel chances that pose new challenges.

In conclusion, TACTICS will follow the NATO recommendations, applying a SOA-based security architecture also supporting attribute-based access control and service delegation with legacy systems on application, services and transport layers. The COI will have to be implemented to ease the management and to investigate signaling optimization regarding security and key distribution, innovative classification and information filtering combination and an innovative methodology for traceability.

Based on this, in addition to label-based classification and filtering mechanisms the identification of information whose aggregation can be sensitive or would require classification is therefore highly relevant and poses a substantial research challenge.

## 6. Conclusions

TACTICS aims at the realization and experimentation of a Tactical Services Infrastructure that enables the participation of tactical radio networks in SOA infrastructures and allows for providing and consuming services to and from the strategic domain, independently of the user's location. We identified the new military capabilities, which most closely correspond to the military user's needs, their enabling requirements, and the underlying key challenges that must be addressed. Then, after relating with the State of the Art work, the progress that TACTICS will bring has been discussed.

## Acknowledgements

This paper describes the work currently undertaken by a Consortium of Companies and Academic Institutes within the B 0980 IAP4 GP TACTICS: "TACTICal Service oriented architecture" project, supported by European Defense Agency.

Tactical Networks", IEEE Military Communications Conference, 2007, pp. 1-7.